\documentclass[11pt,a4paper]{article}
\usepackage{enumerate}
\usepackage{amsmath,amssymb}
\usepackage{amsthm}
\usepackage[breaklinks]{hyperref} 
\usepackage{verbatim}
\usepackage{color}
\usepackage{xcolor}
\usepackage{soul}
\usepackage{bm}
\usepackage{graphicx}
\newcommand{\ifdraft}[1]{}
\definecolor{aocolour}{rgb}{1,0.7,0.8}
\definecolor{iocolour}{rgb}{0.7,0.8,1}

\newtheorem{definition}{Definition} 
\newtheorem{lemma}{Lemma} 
\newtheorem{theorem}{Theorem} 
\newtheorem{example}{Example} 
\newtheorem{claim}{Claim} 

\newcommand{\seepage}[2][see]{\marginpar{\scriptsize (\text{#1} p.~\pageref{#2})}}

\newcommand{\set}[2]{\{\, {#1}\mid{#2}\,\}}

\newcommand{\eps}{\varepsilon}
\renewcommand{\epsilon}{\varepsilon}
\renewcommand{\phi}{\varphi}
\renewcommand{\emptyset}{\varnothing}
\renewcommand{\le}{\leqslant}

\newcommand{\nt}[3]{{}_{#1}#2_{#3}}
\newcommand{\first}[2][k]{\mathrm{First}_{#1}(#2)}

\begin{document}

\sloppy

\title{On LL($k$) linear conjunctive grammars\thanks{%
	This work was supported by
	the Ministry of Science and Higher Education of the Russian Federation,
	agreement 075-15-2019-1619.
	}}
\author{Ilya Olkhovsky\thanks{%
	Department of Mathematics and Computer Science,
	14th Line V.O., 29, Saint Petersburg 199178, Russia
	\emph{and}
	Leonhard Euler International Mathematical Institute at St. Petersburg State University,
	Saint Petersburg, Russia.
	E-mail:
	\texttt{ilianolhin@gmail.com}.}
	\and
	Alexander Okhotin\thanks{%
	Department of Mathematics and Computer Science,
	14th Line V.O., 29, Saint Petersburg 199178, Russia.
	E-mail:
	\texttt{alexander.okhotin@spbu.ru}.}
}

\maketitle

\begin{abstract}
Linear conjunctive grammars are a family of formal grammars
with an explicit conjunction operation allowed in the rules,
which is notable for its computational equivalence fo one-way real-time cellular automata,
also known as trellis automata.
This paper investigates the LL($k$) subclass of linear conjunctive grammars,
defined by analogy with the classical LL($k$) grammars:
these are grammars that admit top-down linear-time parsing with $k$-symbol lookahead.
Two results are presented.
First, every LL($k$) linear conjunctive grammar
can be transformed to an LL(1) linear conjunctive grammar,
and, accordingly, the hierarchy with respect to $k$ collapses.
Secondly, a parser for these grammars
that works in linear time and uses logarithmic space
is constructed,
showing that the family of LL($k$) linear conjunctive languages
is contained in the complexity class $L$.

\end{abstract}

\section{Introduction}

LL($k$) parsing is perhaps the best known linear-time parsing method.
An LL($k$) parser reconstructs a parse tree of the input string top-down,
as it reads the string from left to right.
At each step, the parser selects a rule to apply to a nonterminal symbol,
looking ahead by at most $k$ symbols.

The LL($k$) parsing is applicable
to a subclass of formal grammars known as the \emph{LL($k$) grammars};
the main theoretical properties of these grammars
have been established in the papers 
of Knuth~\cite{Knuth_LL},
Lewis and Stearns~\cite{LewisStearns},
and Rosenkrantz and Stearns~\cite{RosenkrantzStearns}.
In particular, Rosenkrantz and Stearns~\cite{RosenkrantzStearns}
and Kurki-Suonio~\cite{Kurkisuonio} proved that,
for each $k$, the LL($k+1$) grammars can define more languages
than the LL($k$) grammars,
leading to a strict hierarchy of LL($k$) languages by $k$.
     
A natural subclass of \emph{LL($k$) linear grammars},
which obey the LL($k$) restriction
and allow at most one nonterminal symbol on the right-hand side of any rule,
was first studied by Ibarra et al.~\cite{IbarraJiangRavikumar}
and by Holzer and Lange~\cite{HolzerLange},
who have characterized the computational complexity of the languages defined by these grammars.
The language-theoretic properties of linear LL($k$) languages
were recently investigated by Jir\'askov\'a and Kl\'{\i}ma~\cite{JiraskovaKlima}.
Lately, the authors~\cite{linear_CF_LL}
have demonstrated that in the case of LL($k$) linear grammars,
the hierarchy by $k$ collapses,
that is, every language defined by an LL($k$) linear grammar for some $k$
can be defined by an LL(1) linear grammar.
This transformation incurs an exponential blow-up
in the size of the grammar,
and, furthermore, it was proved that this blow-up
is unavoidable in the worst case~\cite{linear_CF_LL}.

The idea of LL($k$) parsing is applicable to several generalizations
of ordinary (``context-free'') formal grammars.
One of such extensions are \emph{conjunctive grammars},
introduced by Okhotin~\cite{Conjunctive},
which enrich the expressive power of ordinary grammars
by allowing a conjunction operation in the rules;
a rule $A \to \alpha \& \beta$ defines all strings
that can be represented both as $\alpha$ and as $\beta$.
The subclass of \emph{LL($k$) conjunctive grammars}
and the associated linear-time parsing algorithm
were defined~\cite{ConjunctiveLL,BooleanLL},
but almost nothing is known about its theoretical properties.

This paper investigates LL($k$) parsing
for a subclass of conjunctive grammars
called \emph{linear conjunctive grammars},
that is, grammars in which every conjunct in every rule may contain at most one nonterminal symbol.
Linear conjunctive grammars
are important for being equivalent to \emph{one-way real-time cellular automata},
also known as \emph{trellis automata}~\cite{LinearAutomata,LinearNonterminals},
and the associated family of languages
has received quite a lot of attention in the literature~\cite{CulikGruskaSalomaa,Dyer,IbarraKim,Terrier},
including some recent work on their expressive power~\cite{Terrier_slender,Terrier_limits}.
Turning to the LL($k$) subfamily of linear conjunctive grammars,
it was proved that they cannot define a language as simple as
$\set{a^n b^n s}{n \geqslant 0, \: s \in \{a,b\}}$~\cite{BooleanLLfamily},
but this is about all that is known about this family.
However, in spite of these grammars' inability to define some particular examples,
this family may still contain some computationally hard specimens.
Furthermore, it remains unknown whether these languages form a hierarchy by $k$.

This paper addresses both the computational complexity of LL($k$) linear conjunctive grammars,
and the existence of a hierarchy by $k$.
First, it is shown that, like for ordinary LL($k$) linear grammars,
the hierarchy by $k$ collapses,
and LL(1) linear conjunctive grammars
are as powerful as LL($k$) linear conjunctive, for any $k$.
Secondly, a parsing algorithm for LL($k$) linear conjunctive grammars is constructed,
which not only works in linear time, but also uses logarithmic space.
Accordingly, all languages defined by LL($k$) linear conjunctive grammars
lie in the complexity class $L$,
and therefore, under the standard assumption that $\mathrm{L} \neq \mathrm{P}$,
these grammars cannot define any P-complete languages,
unlike linear conjunctive grammars without the LL($k$) condition~\cite{IbarraKim}.

\section{Definitions}\label{definitions}

\begin{definition}
A \textbf{linear conjunctive grammar} is a quadruple
$G=(\Sigma, N, R, S)$ that consists of the following components:
\begin{enumerate}
\item $\Sigma$ is the alphabet of the language being defined.
\item $N$ is a finite set of \textbf{nonterminals}. Each nonterminal specifies some property that a given string from $\Sigma^*$ can have or not have.
\item
$R$ is a finite set of \textbf{rules}, with each rule describing a possible structure of a string with a property $A \in N$.
Each rule is either of the form $A\to u_1B_1v_1\;\&\;u_2B_2v_2\;\&\;\cdots\;\&\;u_rB_rv_r$, with $B_1,\ldots,B_r\in N$ and $u_1,v_1,\ldots,u_r,v_r\in\Sigma^*$, or of the form $A\to y$, with $y\in\Sigma^*$.
	
\item $S\in N$ is the \textbf{initial} nonterminal symbol.
\end{enumerate}
\end{definition}

In a rule $A\to u_1B_1v_1\;\&\;u_2B_2v_2\;\&\;\cdots\;\&\;u_rB_rv_r$,
each string $u_i B_i v_i$ is called a \emph{conjunct}.
Such a rule intuitively means that if a string $w$ is representable as each conjunct---or, to be precise, $w=u_is_iv_i$,
where $s_i$ has the property $B_i$, for each $i=1,\ldots,r$---then the string $w$ has the property $A$.
A rule of the form $A\to x$, where $x\in\Sigma^*$,
naturally means that $x$ has the property $A$.

The language described by a conjunctive grammar can be naturally defined by generalizing \textit{parse trees}
used for ordinary (``context-free'') grammars.
This generalization allows leaves to have multiple incoming edges, which correspond to representations of the same substring by different conjuncts.

Parse trees for conjunctive grammars
are just like parse trees in ordinary (``context-free'') grammars,
but whenever a rule involving conjunction is used at a node,
this node has a separate subtrees for each conjunct,
and all these subtrees share the same set of leaves.
Accordingly, these are, strictly speaking,
directed acyclic graphs rather than trees,
but only leaves may have multiple incoming edges.

\begin{figure}[t]
	\centering
	\includegraphics[scale=1]{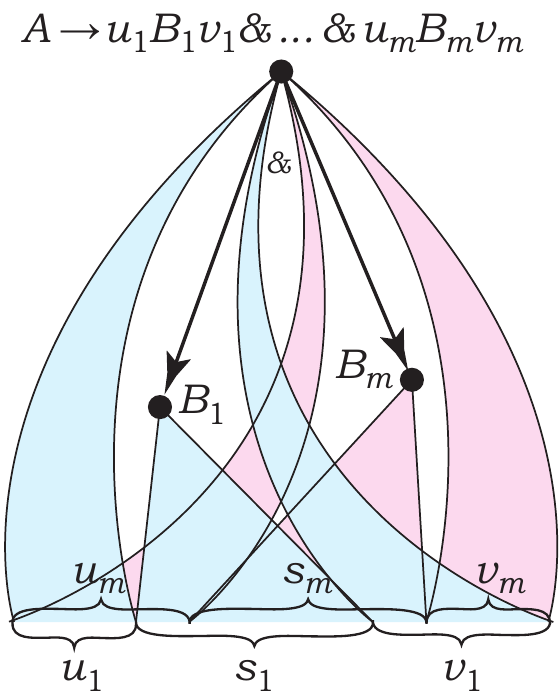}
	\caption{A parse tree of $w$ as $A$}
	\label{f:LinConj_parse_tree_joined}
\end{figure}

\begin{definition}
Let $G=(\Sigma, N, R, S)$ be a linear conjunctive grammar.
A parse tree of a string $w = a_1 \ldots a_n \in \Sigma^*$ as $A \in N$
has $n$ ordered leaves labelled with $a_1, \ldots, a_n$,
a root node labelled with $A$,
and
\begin{itemize}
\item
	either there is a rule $A \to w \in R$,
	and the rule node $A$ has all leaves as its immediate descendants,
\item
	or there exists a rule
	$A \to u_1 B_1 v_1 \,\&\, \ldots \,\&\, u_m B_m v_m$,
	such that, for each $i$-th conjunct, $w=u_i y_i v_i$ for some $y_i$,
	and $A$ has $m$ groups of descendants corresponding to its conjuncts,
	with the group corresponding to each $u_i B_i v_i$
	containing $|u_i B_i v_i|$ immediate descendants:
	the first $|u_i|$ leaves of $w$,
	a node labelled with $B_i$ spanning over the substring $y_i$,
	and the last $|v_i|$ leaves of $w$,
	and, furthermore, the subtree of $B_i$ is a parse tree of $y_i$ as $B_i$.
	
	Figure~\ref{f:LinConj_parse_tree_joined} illustrates groups of descendants of a node $A$.
\end{itemize}
A parse tree of a string as $S$ is called simply a parse tree.

Each nonterminal $A$ defines a language $L_G(A)=L(A)$,
which is the set of all strings $w$, for which there exists a tree of $w$ as $A$.
The language defined by the grammar is the language defined by its initial symbol: $L(G)=L_G(S)$.

The language defined by a conjunction
$\phi=\alpha_1\,\&\,\ldots\,\&\,\alpha_r$
is defined as
$L(\phi)=L(\alpha_1)\cap\ldots\cap L(\alpha_r)$.
\end{definition}

\begin{figure}[t]
	\centering
	\includegraphics[scale=1]{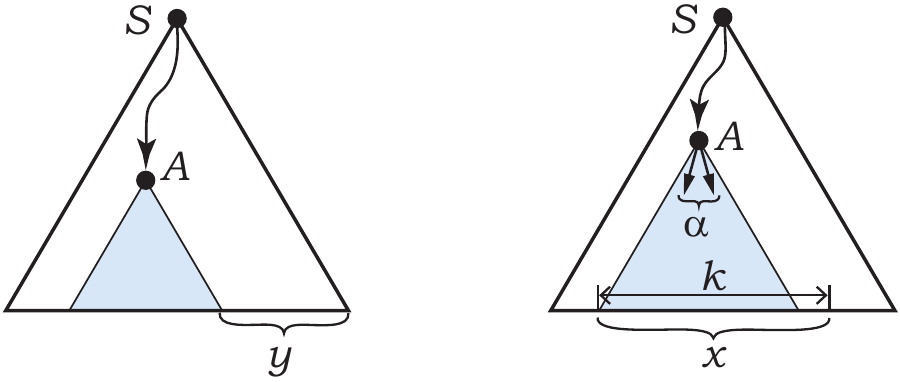}
	\caption{String $y$ follows an $A$-subtree (left) and string $x$ defines the rule $\alpha$ (right)}
	\label{f:definitions}
\end{figure}

\begin{definition}
Let $G=(\Sigma, N, R, S)$ be a linear conjunctive grammar,
and let $\tau$ be a subtree of some parse tree,
with the root of $\tau$ labelled with $A \in N$ (an \emph{$A$-subtree}).
Let $y$ be the string of all leaves located to the right of the rightmost leaf in $\tau$.
Then it is said that $y$ \emph{follows} the subtree $\tau$. This is illustrated in Figure~\ref{f:definitions}(left).
\end{definition}

The class of LL($k$) linear conjunctive grammars studied in this paper
is defined by the following restriction.

\begin{definition}
An LL($k$)-table for a linear conjunctive grammar $G=(\Sigma,N,R,S)$ is a partial function $T \colon N\times\Sigma^{\leqslant k}\to R$,
which satisfies the following condition.
For every subtree $\tau$ of any parse tree, let $A \in N$ be the label of the root of $\tau$, and let $x \in \Sigma^{\leqslant k}$ be the first $k$ leaves starting from the first leaf of $\tau$;
then, the rule applied to the root of $\tau$ must be $T(A,x)$,
as illustrated in Figure~\ref{f:definitions}(right).

If an LL($k$) table for a grammar $G$ exists, then $G$ is said to be LL($k$).
\end{definition}

\begin{example}\label{an_bn_cn_LinConjLL_example}
The following LL($k$) linear conjunctive grammar
defines the language $\set{a^n b^n c^n}{n \geqslant 0}$.
\begin{equation*}\begin{array}{rcl}
	S &\to& A \,\&\, C \\
	A &\to& aA \ | \ D \\
	D &\to& bDc \ | \ \epsilon \\
	C &\to& aCc \ | \ B \\
	B &\to& bB \ | \ \epsilon
\end{array}\end{equation*}

The LL(1) table for this grammar is given below.
\begin{equation*}
	\begin{array}{|c|cccc|}
	\hline
		& \epsilon
			& a
				& b
					& c \\
	\hline
	S
		& S \to A \& C
			& S \to A \& C
				& -
					& - \\
	A
		& A \to D
			& A \to a A
				& A \to D
					& - \\
	D
		& D \to \epsilon
			& -
				& D \to b D c
					& D \to \epsilon \\
	C
		& C \to B
			& C \to a C c
				& C \to B
					& - \\
	B
		& B \to \epsilon
			& -
				& B \to b B
					& B \to \epsilon \\
	\hline
	\end{array}
\end{equation*}
\end{example}

\section{The aligned form of an LL(k) linear conjunctive grammar}\label{grammar_aligning_section}

In this section it is shown
that every LL($k$) linear conjunctive grammar
can be transformed to a normal form called the \textit{aligned form},
which is similar to the Greibach normal form for non-linear grammars.

\begin{definition}
A linear conjunctive grammar $G=(\Sigma, N, R, S)$ is called \textbf{aligned},
if each rule in $G$ is either of the form
$A\to a C_1 v_1\,\&\ldots\&\, a C_m v_m$,
with $a \in \Sigma$, $m \geqslant 1$, $C_1, \ldots, C_m \in N$
and $v_1, \ldots, v_m \in \Sigma^*$,
or of the form $A\to y$, with $y \in \Sigma^*$.
\end{definition}

In general, it is likely that some linear conjunctive grammars
cannot be transformed to the aligned form
(although, as to the authors' knowledge, no proof has ever been presented).
However, for the LL($k$) subclass, a transformation turns out to be possible.

The transformation consists of two steps:
first, so called \textit{left-recursive} rules are eliminated from the grammar,
and then each rule $A\to u_1C_1v_1\,\&\ldots \&\,u_m C_m v_m$
is ``aligned'' by introducing new nonterminal symbols,
so that in each conjunct there is exactly one symbol before a nonterminal.

\begin{definition}
A rule $A\to\alpha_1\&\ldots\&\alpha_r$ is called \textbf{left-recursive},
if at least one of its conjuncts
is of the form $\alpha_j = Bt$,
for some $B \in N$ and $t \in \Sigma^*$.
\end{definition}

Before describing the transformation, it is convenient to establish
the uniqueness of parse trees in LL($k$) linear conjunctive grammars,
which will be used many times throughout the paper.
The result holds for all LL($k$) conjunctive grammars, not necessary linear.
However, for simplicity, the proof is given only in the linear case.

\begin{lemma}\label{LL_subtrees_equivalence}
Let $G$ be an LL($k$) linear conjunctive grammar,
and let $\tau_1$ and $\tau_2$ be any two parse trees for a string $w \in \Sigma^*$.
Let $\tau^A_1$ and $\tau^A_2$ be two $A$-subtrees in $\tau_1$ and in $\tau_2$, respectively,
such that the leaves to the left of each subtree form the same string $s \in \Sigma^*$.
Then, $\tau^A_1$ and $\tau^A_2$ are identical,
and, in particular, define the same substring of $w$.
\end{lemma}
\begin{proof}
Since the selected subtrees of $\tau_1$ and of $\tau_2$
are positioned within these trees
after the same string of leaves $s$,
the first $k$ leaves starting from the first leaf of $\tau^A_1$
and the first $k$ leaves starting from the first leaf of $\tau^A_2$
form the same substring $x$.
Then, since the grammar $G$ is LL($k$),
the same rule $T(A, x)$ is applied at the roots of both subtrees.
the rule applied to the root of each subtree $\tau^A_1$ and $\tau^A_2$ is $T(A,x)$.

Now it is claimed that $\tau^A_1$ and $\tau^A_2$ are identical.
This is proved by induction on the height of $\tau^A_1$.

If $\tau^A_1$ consists of a single rule $A\to y$,
n the same rule is applied to the root of $\tau^A_2$,
and thus $\tau^A_2$ also consists of a single rule $A\to y$.
This substring $y$ is the one immediately following $s$ in $w$
(that is, $w\in sy \Sigma^*$).

Now assume that the rule applied to the root of $\tau^A_1$
is $A\to u_1C_1v_1\,\&\ldots\&\, u_mC_mv_m$.
Then the rule applied to the root of $\tau^A_2$
is also $A\to u_1C_1v_1\,\&\ldots\&\, u_mC_mv_m$.
For each $j\in\{1,\ldots,m\}$,
denote by $\tau^j_1$ and $\tau^j_2$
the subtrees corresponding to the conjunct $u_jC_jv_j$ in $\tau^A_1$
and in $\tau^A_2$, respectively.
By the induction hypothesis, for each $j\in\{1,\ldots,m\}$,
the subtrees $\tau^j_1$ and $\tau^j_2$ are identical.
Therefore, the subtrees $\tau^A_1$ and $\tau^A_2$ are also identical.
\end{proof}

The next lemma establishes that it is possible to remove left-recursive rules from each LL($k$) linear conjunctive grammar.

\begin{lemma}\label{left_recursion_elimination_lemma}
For every LL($k$) linear conjunctive grammar $G=(\Sigma, N, R, S)$,
there exists an LL($k$) linear conjunctive grammar $G'=(\Sigma,N,R',S)$
without left-recursive rules
that defines the same language as $G$.
\end{lemma}
\begin{proof}
In the new grammar $G'$,
each rule will simulate a certain fragment of a parse tree in $G$,
comprised of node and a tree of all left-recursive chains
coming out of this node in different conjuncts.

Let $\tau$ be a parse tree in $G$ with a selected $A$-subtree $\tau_A$,
and let $x$ be the string formed by the first $k$ leaves in the tree,
starting from the first leaves of the $A$-subtree.
Denote by $w$ the string defined by $\tau$,
and denote by $y$ the substring of $w$ defined by $\tau_A$.

Every such pair $(\tau,\tau_A)$
defines a rule $A\to\phi(\tau,\tau_A)$ in the new grammar.
It will be shown later that the resulting set of rules is finite.

Let us call a conjunct $\alpha$ \emph{normal} if it is not left-recursive,
that is, either $\alpha=uCv$ for some $C\in N$ and $u\in\Sigma^+,v\in\Sigma^*$,
or $\alpha \in \Sigma^*$.

A \textit{left chain} is a path $v_0\to v_1\to\ldots\to v_m$ in a parse tree,
wherein a left-recursive rule $B_j\to\ldots\&\,B_{j+1}t_{j+1}\,\&\ldots$
is applied to each vertex $v_j$ with $j<m$,
and $v_{j+1}$ is the immediate descendant of $v_j$
corresponding to the conjunct $B_{j+1}t_{j+1}$.

Denote by $v_A$ the root of the $A$-subtree.
A normal conjunct $\alpha$ \textit{is reachable from $A$ via a left chain},
if there exists a left chain $v_0=v_A\to v_1\to\ldots\to v_m$,
wherein the rule applied to $v_m$
is $B_m\to\ldots\&\,\alpha\,\&\ldots$,
as in Figure~\ref{f:left_chain}. 

\begin{figure}[t]
	\centering
	\includegraphics[scale=1]{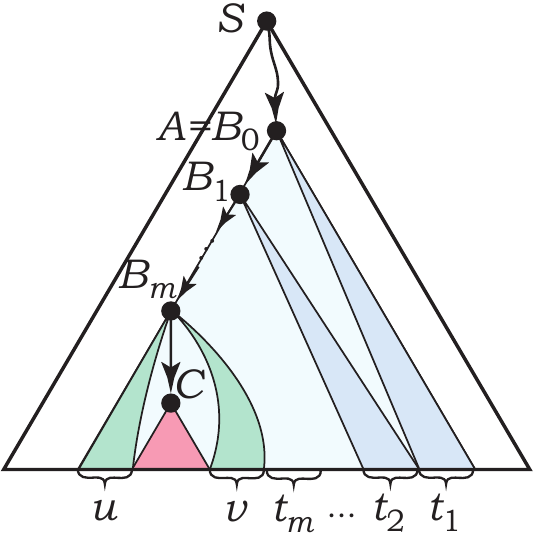}
	\caption{Reachability of a normal conjunct $\alpha=uCv$ from $A$ by a left chain.}
	\label{f:left_chain}
\end{figure}

For convenience of the further proof,
let us define \textit{left chains isomorphism}.

\begin{definition}
Let $\tau$ and $\tau'$ be two parse trees in $G$,
and let $v_0\to\ldots\to v_m$ and $v_0'\to\ldots\to v_m'$
be left chains in $\tau$ and in $\tau'$, respectively.

The left chains $v_0\to\ldots\to v_m$ and $v_0'\to\ldots\to v_m'$
are called \textit{isomorphic},
if, for each $j\in\{0,\ldots,m\}$,
the following conditions hold:
\begin{itemize}
\item The vertices $v_j$ and $v_j'$ are labelled with the same nonterminal.
\item
For $j < m$, if the rule applied to $v_j$ is $B_j\to \ldots\&\, B_{j+1}t_{j+1}\,\&\ldots$,
where the conjunct $B_{j+1}t_{j+1}$ corresponds to the vertex $v_{j+1}$,
then the same rule is applied to $v_j'$ in the other chain,
and the immediate descendant of $v_j'$
corresponding to the conjunct $B_{j+1}t_{j+1}$ is $v_{j+1}'$.
\end{itemize}
\end{definition}

The next claim establishes that
the set of all left chains beginning in some vertex $v$
(and hence also the set of all normal conjuncts reachable from $v$ via left chains)
is, up to isomorphism, defined just by the nonterminal at $v$
and by the first $k$ leaves
starting from the first leaf of the subtree of $v$,
and does not depend on the rest of the parse tree.

\begin{claim}\label{normal_conjuncts_set_depends_only_on_x}
Let $\tau$ and $\tau'$ be two parse trees in $G$
with selected $A$-subtrees $\tau_A$ and $\tau_A'$, respectively.
Assume that in both parse trees,
the first $k$ leaves, starting from the first leaves of $A$-subtrees,
form the same string $x$.

Then, for each left chain $v_0=v_A\to v_1\to\ldots\to v_m$ in $\tau$,
there exists an isomorphic left chain $v_0'=v_A'\to v_1'\to\ldots\to v_m'$ in $\tau'$.
\end{claim}
\begin{proof}
The claim is proved by induction on $m$, the length of a left chain in $\tau$.

For $m=0$,
both left chains consist of a single vertex $A$, which makes them isomorphic.

Now assume that $m>0$,
and let $v_0=v_A\to v_1\to\ldots\to v_m$ be a left chain in $\tau$,
wherein some rule $B_j\to\ldots\&\, B_{j+1}t_{j+1}\,\&\ldots$
is applied to each vertex $v_j$, with $j<m$.

By the induction hypothesis,
there exists a left chain $v_0'\to v_1'\to\ldots\to v_{m-1}'$ in $\tau'$,
which is isomorphic to $v_0\to v_1\to\ldots\to v_{m-1}$.
Since each conjunct $B_jt_j$ begins with a nonterminal symbol,
the subtrees of $v_{m-1}$ and $v_{m-1}'$
have the same first leaf, and hence the same string of first $k$ leaves
beginning at this position.
Therefore,  the same rule $B_{m-1}\to \ldots\&\, B_m t_m \,\&\ldots$
is applied to both subtrees $v_{m-1}$ and $v_{m-1}'$.

Then $v_m'$ can be defined as the immediate descendant of $v_{m-1}'$
corresponding to the conjunct $B_mt_m$,
and this makes the left chain $v_0'\to v_1'\to\ldots\to v_m'$
isomorphic to $v_0\to v_1\to\ldots\to v_m$.
\end{proof}

A normal conjunct can be reachable from $A$
via several different left chains,
and thus each normal conjunct $uCv$
can correspond to several different subtrees in $\tau_A$.
The next claim establishes that all such subtrees are identical.

\begin{claim}
Let $\alpha=uCv$ be a normal conjunct
reachable from $v_A$ via two left chains,
$v^1_0=v_A\to\ldots\to v^1_{m_1}$ and $v^2_0=v_A\to\ldots\to v^2_{m_2}$,
and let $\tau^1_C$ and $\tau^2_C$ be the subtrees
corresponding to $\alpha$ in these left chains.
Then the subtrees $\tau^1_C$ and $\tau^2_C$ are identical.
\end{claim}
\begin{proof}
The roots of the subtrees $\tau^1_C$ and $\tau^2_C$
are both labelled with the same nonterminal $C$,
and the leaves to the left of each subtree
form the same string $su$,
where $s$ is the string of all leaves before $\tau_A$.
Then, by Lemma~\ref{LL_subtrees_equivalence},
the subtrees $\tau^1_C$ and $\tau^2_C$ are identical.
\end{proof}

Let $\alpha$ be a normal conjunct reachable from $A$ via some left chain.
Denote by $t_\alpha$ the string of all leaves of $\tau_A$
following the subtree corresponding to $\alpha$.

Now the rule $A\to\phi(\tau,\tau_A)$ can be defined.

Assume that there exists a normal conjunct without nonterminals,
which is reachable from $A$ via a left chain.
Then, fix any such conjunct and denote it by $\sigma\in\Sigma^*$.
The rule $A\to \phi(\tau,\tau_A)$ is then defined as $A\to\sigma t_{\sigma}$.
Note that in this case $\sigma t_\sigma=y$,
and hence the rule does not depend on the choice of the conjunct $\sigma$.
However, one has to fix some conjunct $\sigma$,
because the correctness proof
requires a partition of the rule in the form $A\to\sigma t_\sigma$.

Otherwise, let $\alpha_1$,\ldots,$\alpha_r$ be all the conjuncts
reachable from $A$ via left chains.
Each $\alpha_j$ contains a nonterminal.
The right-hand side $\phi(\tau,\tau_A)$ is then defined
as the conjunction $\alpha_1 t_{\alpha_1}\&\ldots\&\alpha_r t_{\alpha_r}$.

Accordingly, the rule $A\to\phi(\tau,\tau_A)$ is defined
by the set of all left chains $v_0\to\ldots\to v_m$,
by which normal conjuncts are reachable from $A$.
The latter set, on the other hand,
is defined by the string $x$
by Claim~\ref{normal_conjuncts_set_depends_only_on_x}.

Therefore, the resulting rule $A\to\phi(\tau,\tau_A)$ is also defined by $x$,
and does not depend on the rest of $\tau$.
In particular, the set of all rules $A\to\phi(\tau,\tau_A)$
constructed for all possible pairs $(\tau,\tau_A)$ is finite.

Now the new grammar $G'=(\Sigma,N,R',S)$ is defined as follows.
The set of nonterminals and the initial nonterminal of $G'$
are the same as those in $G$,
and the set of rules $R'$ consists of all rules
constructed for each possible pair $(\tau,\tau_A)$ and for each nonterminal $A\in N$.

By construction, each rule of the new grammar $G'$
is of the form $A\to\alpha_1 t_{\alpha_1}\&\ldots\&\alpha_r t_{\alpha_r}$.
Let us fix such a partition for each rule;
if the same rule can be obtained from different pairs $(\tau,\tau_A)$,
then choose a partition corresponding to any pair.

By the construction, $G'$ is linear
and does not contain any left-recursive rules.

The proof that $G'$ is LL($k$) and defines the same language as $G$
is based on a one-to-one correspondence between parse trees in $G$ and $G'$.

The following claim establishes
how a parse tree in the new grammar
can be obtained from a parse tree in the original grammar.
\begin{claim}\label{h_map}
Every string defined in $G$ is also defined in $G'$.
\end{claim}
\begin{proof}
Let $w$ be any string in $L(G)$ and fix its parse tree $\widehat{\tau}$.
It is claimed that,
for every $A$-subtree $\tau$ of $\widehat{\tau}$,
which defines some substring $y$,
the string $y$ is defined by $A$ in $G'$.
The proof is given by induction on the height of $\tau$.

Assume that in $\tau$
there exists a normal conjunct $\sigma\in\Sigma^*$
reachable from $A$ via a left chain,
and let $t_\sigma$ be the leaves of $\tau$ following $\sigma$,
so that $\tau$ defines the string $\sigma t_\sigma = y$.
Then, by construction, $G'$ contains a rule $A\to \sigma t_\sigma$,
and accordingly $y \in L_{G'}(A)$, as claimed.
In particular, this argument covers of $\tau$ of minimal height,
when it consists of a single $A\to\sigma$ with $\sigma\in\Sigma^*$,
proving the the base case of induction.

Now let $u_1C_1v_1$, \ldots, $u_rC_rv_r$
be all normal conjuncts reachable from $A$ via left chains,
and let $t_1$, \ldots, $t_r$ be the ``tails''
corresponding to these conjuncts,
so that for each $j\in\{1,\ldots,r\}$,
the $C_j$-subtree in $\tau$ is followed by the string $v_jt_j$.
Denote by $\tau_j$ the $C_j$-subtree in $\tau$
corresponding to the conjunct $u_jC_jv_j$. 
Let $z_j$ be the substring defined in $\tau_j$.
By the induction hypothesis, the $z_j$ is defined by $C_j$ in $G'$.

For each $j$-th conjunct, $u_j z_j v_j t_j = y$.
By the construction, $G'$ contains a rule $A\to u_1C_1v_1t_1\,\&\ldots\&\,u_rC_rv_rt_r$.
Then, $y \in L_{G'}(A)$ by this rule.
\end{proof}

Similarly, from each parse tree in the new grammar,
one can construct a parse tree in the original grammar.
In doing so, the correspondence between the vertices of the parse tree in the new grammar
nd the vertices of the parse tree in the original grammar is defined.
In this correspondence, each vertex of the parse tree in the new grammar
is mapped to a vertex of the parse tree in the original grammar,
but some vertices of the parse tree in the original grammar
do not occur as an image of any vertex in the parse tree in the new grammar.

\begin{claim}\label{h'_map}
There exists a function $h' \colon \tau' \mapsto (\tau, \rho)$,
which maps a parse tree $\tau'$ of a string as a nonterminal $D$ in $G'$
to a parse tree $\tau$ of the same string as $D$ in $G$
and to a mapping $\rho$ from the set of vertices of $\tau'$
to the set of vertices of $\tau$,
such that:
\begin{enumerate}
\item $\rho$ maps the root of $\tau'$ to the root of $\tau$.
\item If a vertex $v$ is labelled with a nonterminal $A$,
then the vertex $\rho(v)$ is also labelled with $A$.
\item The subtree of $\rho(v)$ defines the same string as the subtree of $v$.
\item The subtree of $\rho(v)$ is followed by the same string as the subtree of $v$.
\item   
    \begin{enumerate}
    \item 
    Suppose that the rule applied to $v$ is $A\to y$,
    and the partition corresponding to that rule
    is $y=\sigma t_\sigma$.
    Then the conjunct $\sigma$
    is reachable from $\rho(v)$ via a left chain,
    and the leaves of the subtree of $\rho(v)$ following this conjunct
    form the string $t_\sigma$.
\label{rho_property_short}
    \item 
    Suppose that the rule applied to $v$ is
    $A\to u_1C_1z_1\,\&\ldots\,\& u_r C_rz_r$,
    where the partition fixed for each $z_j$ is $z_j=v_j t_j$.

    Then the set of all normal conjuncts
    reachable from $\rho(v)$ via left chains
    is $\{u_1 C_1 v_1,\ldots, u_rC_rv_r\}$,
    and each $C_j$-subtree in the subtree of $\rho(v)$, with $j\in\{1,\ldots,r\}$,
    is followed by string $v_jt_j$.
    \label{rho_property_long}
    \end{enumerate}
\end{enumerate}
\end{claim}

The function $h'$ maps each parse tree $\tau'$ to a pair $(\tau,\rho)$.
However, in the following, for simplicity,
$h'$ will be used as if it maps parse trees to parse trees,
and the mapping $\rho$ exists separately from $h'$.

\begin{proof}
The proof is carried out by induction on the height of $\tau'$.
Let $A$ be the label of the root of $\tau'$.

In the base case, $\tau'$ consists of a single rule $A\to y$, with $y\in\Sigma^*$.
By the construction of $G'$,
there is a partition $y=\sigma t_\sigma$,
such that there exists a parse tree in $G$,
with an $A$-subtree,
wherein the normal conjunct $\sigma$ is reachable from $A$,
and the leaves to the right of $\sigma$ form the string $t_\sigma$.
Denote that $A$-subtree by $\tau$,
and define $h'(\tau')=\tau$.
The function $\rho$ maps the root of $\tau'$ to the root of $\tau$. 

Now assume that the rule applied to the root of $\tau'$
is $A\to u_1C_1z_1 \;\&\; \ldots \;\&\; u_rC_rz_r$,
wherein each $z_j$ is partitioned as $z_j=v_jt_j$
according to the construction.
For each $j\in\{1,\ldots,r\}$,
let $\tau_j'$ denote the $C_j$-subtree of $\tau'$
corresponding to the conjunct $u_jC_jz_j$.
By the induction hypothesis, for each $j\in\{1,\ldots,r\}$
there is a parse tree $h'(\tau_j')$,
which has the same root and defines the same string as $\tau_j'$,
as well as a mapping $\rho_j$ from the set of vertices of $\tau_j'$
to the set of vertices of $h'(\tau_j')$
satisfying the condition in Claim~\ref{h'_map}.

By construction,
there exists a parse tree in $G$
with an $A$-subtree $\tau_A$,
such that the set of all normal conjuncts reachable from $A$ via left chains
is $\{u_1C_1v_1, \ldots, u_rC_rv_r\}$,
and for each $j\in\{1,\ldots,r\}$,
the string following the conjunct $u_jC_jv_j$ in $\tau_A$ is $t_j$. 

Let $\tau_j$ be the subtree in $\tau_A$
corresponding to the conjunct $u_jC_jv_j$.
Then the parse tree $h'(\tau')$ is obtained from $\tau_A$
by replacing each subtree $\tau_j$ with the subtree $h'(\tau_j')$.

The mapping $\rho$ for $\tau'$ is defined as follows:
the root of $\tau'$ is mapped to the root of $h'(\tau')$,
and vertices from each subtree $\tau_j'$
are mapped to the corresponding vertices of $h'(\tau_j')$
by the mapping $\rho_j$.
\end{proof}
The last claim immediately entails $L(G')\subseteq L(G)$,
and therefore the equality $L(G)=L(G')$ is proved.

Consider any vertex $v$ in some parse tree $\tau'$ in the new grammar.
The rule $A\to \phi(\tau,\tau_A)$ applied to $v$
is obtained from some parse tree $\tau$ in the original grammar,
with a selected $A$-subtree $\tau_A$.
The function $h'$ from claim~\ref{h'_map}, on the other hand,
matches $\tau'$ to some parse tree $h'(\tau')$,
and matches the subtree in $\tau'$ with the root $v$
to a subtree in $h'(\tau')$ with the root $\rho(v)$.

The next claim states that the rule $A\to \phi(\tau,\tau_A)$
applied at a vertex $v$
coincides with the rule obtained from the parse tree $h'(\tau')$
with the selected subtree $\rho(v)$.

\begin{claim}\label{rule_construction_commutes_with_h'}
Let $\tau'$ be a parse tree in $G'$,
let $v$ be a vertex in $\tau'$,
and let $A\to\psi$ be the rule applied at $v$.
Then $\psi=\phi(h'(\tau'),\tau_{\rho(v)})$,
where $\tau_{\rho(v)}$ is the subtree of $h'(\tau')$ with the root $\rho(v)$.
\end{claim}
\begin{proof}
First consider the case of $\psi=y$, with $y\in\Sigma^*$.
By the construction of $G'$,
there is a partition $y=\sigma t_\sigma$ fixed for $y$.
By Claim~\ref{rho_property_short}, the conjunct $\sigma$
is reachable from $\rho(v)$ via a left chain,
and the subtree $\rho(v)$ defines the string $y=\sigma t_\sigma$.
Then, by the construction, $\phi(h'(\tau'),\tau_{\rho(v)})=y=\psi$.

Now assume that $\psi=u_1C_1z_1 \;\&\; \ldots \;\&\; u_rC_rz_r$,
and the partition fixed for each $z_j$ is $z_j=v_jt_j$.
Then, by Claim~\ref{rho_property_long},
the set of all normal conjuncts reachable from $\rho(v)$ via left chains
equals $\{u_1 C_1 v_1,\ldots, u_rC_rv_r\}$,
and for each $j\in\{1,\ldots,r\}$,
the subtree corresponding to $u_jC_jv_j$
is followed in the subtree of $\rho(v)$ with with string $v_jt_j$.
Therefore, by the construction,
$\phi(h'(\tau'),\tau_{\rho(v)})=u_1C_1v_1 t_1 \;\&\; \ldots \;\&\; u_rC_rv_r t_r=\psi$.
\end{proof}

It remains to prove that $G'$ is LL($k$).

\begin{claim}
The grammar $G'$ is LL($k$).
\end{claim}
\begin{proof}
Let $\tau_1'$ and $\tau_2'$ be two parse trees in $G'$,
each containing an $A$-subtree,
and sharing the same substring $a$ forming the first $k$ leaves,
starting with the first leaves of $A$-subtrees in $\tau_1'$ and $\tau_2'$.

It is claimed that the rules applied to the roots of the $A$-subtrees are the same.

Let $\tau_1=h'(\tau_1')$ and $\tau_2=h'(\tau_2')$.
Let $v_1$ and $v_2$ be the vertices corresponding to the $A$-subtrees
in $\tau_1'$ and $\tau_2'$, respectively.
Let the rule applied to $v_1$ in $\tau_1'$ be $A\to \phi_1$,
and let the rule applied to $v_2$ in $\tau_2'$ be $A\to \phi_2$. 
By Claim~\ref{h'_map}, in both parse trees $\tau_1$ and $\tau_2$,
the first $k$ leaves starting from the first leaves of subtrees of $\rho(v_1)$ and $\rho(v_2)$, respectively, form the same string $x$.
Then, by the construction of rules in the new grammar, $\phi(h'(\tau_1'),\tau_{\rho(v_1)})=\phi(h'(\tau_1'),\tau_{\rho(v_2)})$.

On the other hand, Claim~\ref{rule_construction_commutes_with_h'} entails $\phi_1=\phi(h'(\tau_1'),\tau_{\rho(v_1)})$ and $\phi_2=\phi(h'(\tau_2'),\tau_{\rho(v_2)})$. Therefore, $\phi_1=\phi_2$, and the proof is complete.
\end{proof}
\end{proof}

Once all left-recursive rules are removed from the grammar,
the latter can be made aligned by a direct construction.

\begin{lemma}\label{for_each_grammar_exists_aligned_grammar}
For each LL($k$) linear conjunctive grammar $G=(\Sigma,N,R,S)$,
there exists an aligned LL($k$) grammar $G'$
that defines the same language.
\end{lemma}
\begin{proof}
By Lemma~\ref{left_recursion_elimination_lemma},
it may be assumed that $G$ does not contain left-recursive rules.

If $G$ is not aligned,
then $G$ contains a rule of the form $A\to \ldots\,\&auBv\&\ldots$,
with $|u| \geqslant 1$.
Then a new nonterminal $C$ with a single rule $C\to uBv$ is introduced,
and the rule $A\to \ldots\,\& auBv \&\ldots$ in $G$
is replaced with $A\to \ldots\,\&aC\&\ldots$.
This is repeated until $G$ becomes aligned.

Such a substitution does not affect the language defined by grammar and the LL($k$) property. Thus, the resulting grammar is LL($k$)
and defines the same language as $G$. It is aligned by construction.
\end{proof}

\section{Transforming an LL(k) linear conjunctive grammar to LL(1) linear conjunctive}\label{LL_k_to_LL_1_section}

\begin{theorem}\label{conj_ll_k_to_ll_1_theorem}
For each LL($k$) linear conjunctive grammar $G=(\Sigma, N, R, S)$,
there exists an aligned LL(1) grammar $G'$ that defines the same language.
\end{theorem}

The proof of the theorem is naturally split into several stages of construction.
The main idea of the construction repeats the idea of construction
in the analogous theorem for ordinary, non-conjunctive LL($k$) linear grammars~\cite[Thm.~1]{linear_CF_LL}:
it splits into the same stages of transformation,
and the actual construction is directly generalized.
Some proofs are different from the non-conjunctive case
only in the use of conjunction,
and are accordingly omitted in this paper.
Other parts of the argument, such as the verification of the LL property,
require a more detailed analysis of parse trees;
there proofs are presented in full.

The main idea of the construction is as follows.
Every LL($k$) conjunctive grammar $G=(\Sigma, N, R, S)$
can be implemented in an LL($k$)-parser,
which reads the input string symbol by symbol from left to right,
and attempts to construct its parse tree along with reading it~\cite{ConjunctiveLL}.
The parse tree is constructed top-down.
At each step, the LL($k$) parser has the next $k$ input symbols available,
and for each unprocessed node in the parse tree,
it determines the rule to apply to the nonterminal symbol in this node
by accessing the LL-table, indexed by the nonterminal symbols and the $k$ look-ahead symbols.
The unprocessed nodes of the parse trees,
whose subtrees have not been constructed yet,
are stored in a so-called \emph{tree-structured stack};
but these details of the general algorithm are beyond the scope of this paper.

The task is to reconstruct a given grammar $G$
to obtain an LL(1) linear conjunctive grammar $G'$
that defines the same language as $G$.
A hypothetical LL(1)-parser for a grammar $G'$
should select a rule to use at a node of the parse trees,
using only a single next input symbol.
The main idea is to let the LL(1)-parser \emph{delay the choice of a rule}
until it reads all $k$ next input symbols,
which uniquely determine the rule in the original grammar $G$.

This is done by attaching a \emph{buffer} of at most $k-1$ symbols
to each nonterminal $A$ of the original grammars.
Accordingly, the nonterminals in $G'$ are of the form $\nt{u}{A}{}$,
where $A\in N$ and $u\in\Sigma^{\le k-1}$.
Until the buffer of a nonterminal $\nt{u}{A}{}$ is not yet filled,
the parser applies the following rules for filling up the buffer.
\begin{align*}
	\nt{u}{A}{} \to a\,\nt{ua}{A}{}		&& (|u|<k-1)
\end{align*}

Once the buffer of $\nt{u}{A}{}$ is filled, that is, $|u|=k-1$,
the $k-1$ symbols of the buffer, together with the next input symbol $a$
available to the LL(1)-parser, together form the $k$ symbols
necessary to determine the rule $T(A,ua)$,
which should be applied to $A$ in the original grammar.
At the same time, one should somehow remove the previusly read substring $u$
from the rule $T(A,ua)$,
and this may cause problems if the rule $T(A,ua)$ is ``short'',
that is, if $T(A,ua)$ is of the form $A\to y$, where $y\in\Sigma^*$ and $|y|<|u|$.

In order to avoid this problematic case,
all such ``short'' rules are to be removed from the grammar $G$ beforehand.
Thus, the entire construction consists of two stages:
first, the short rules are removed from the grammar,
and then, using the resulting grammar free of short rules,
an LL(1) linear conjunctive grammar is constructed
using the above idea of buffering lookahead symbols in the nonterminal's subscript.

The elimination of short rules does not use the LL($k$) property of the grammar,
and can be carried out for every linear conjunctive grammar.
The transformation of a grammar without short rules to LL(1)
in turn does not rely on the linearity of the grammar,
and can be done for every LL($k$) conjunctive grammar without short rules.
Nevertheless, for simplicity, at each stage
the grammar is assumed to be both linear and LL($k$).
Furthermore, by Lemma~\ref{for_each_grammar_exists_aligned_grammar},
the original grammar $G$ can be assumed to be aligned.

\subsection{Short rules elimination}\label{section_conj_short}

\begin{figure}[t]
	\centering
	\includegraphics[scale=1]{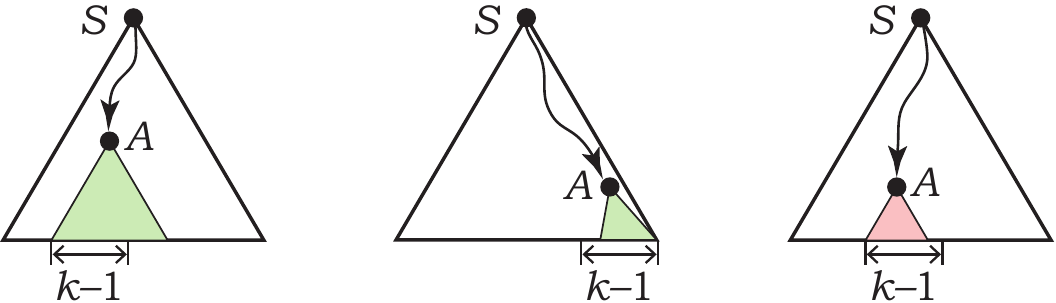}
	\caption{The third rule is short, while the first two are not.}
	\label{f:short_rules}
\end{figure}

\begin{definition}
A rule $A \to y$ is called \textbf{short} if $|y| < k-1$
and there exists a parse tree with an $A$-subtree followed
by a nonempty string,
as in Figure~\ref{f:short_rules} (right).
\end{definition}

\begin{lemma}\label{conj_short_rule_elimination_lemma}
For each aligned LL($k$) grammar $G=(\Sigma, N, R, S)$,
there exists an aligned LL($k$) grammar $G'$
without short rules that defines the same language.
\end{lemma}
\begin{proof}
Nonterminals in the new grammar $G'=(\Sigma, N', R', \nt{}{S}{\eps})$
are of the form $\nt{}{A}{u}$,
with $A \in N$ and $u \in \Sigma^{\leqslant k-1}$.
The intention is to have $L_{G'}(\nt{}{A}{u}=\set{yu}{y \in L_G(A)}$.

Each rule for a nonterminal $\nt{}{A}{u}\in N'$
is obtained by appending the suffix $u$ to the right-hand side
of some rule of the original grammar.

For each rule $A\to y$ in the original grammar,
the new grammar has a rule with the suffix $u$ appended.
\begin{equation*}
	\nt{}{A}{u} \to yu
\end{equation*}

For each rule $A\to aB^1v_1\,\&\ldots\&\,aB^mv_m$ in the original grammar,
the new grammar has a rule
\begin{equation*}
    \nt{}{A}{u}\to a\,\nt{}{B^1}{s_1}\, t_1\,\&\ldots\&\, a\,\nt{}{B^m}{s_m}\, t_m
\end{equation*}
wherein for each $j\in\{1,\ldots,m\}$,
the conjunct $a\,\nt{}{B^j}{s_j} t_j$ is obtained from the conjunct $a B^j v_j$ as follows.
The string $s_j$ consists of the first $k-1$ symbols of the string $v_ju$
(if $|v_ju|\leqslant k-1$ then $s=v_ju$),
and the string $t_j$ consists of the remaining suffix of $v_ju$,
so that $s_j$ and $t_j$ satisfy $s_j=\first{v_ju}$ and $s_jt_j=v_ju$.
The intuition behind this
is that string $u$ is first appended to the conjunct $aB^jv_j$,
and then the longest possible prefix of $v_ju$ is moved to the subscript of $B^j$.
\begin{equation*}
    aB^jv_j\Rightarrow aB^jv_ju\Rightarrow a\,\nt{}{B^j}{s_j}\, t_j
\end{equation*}

The proof of correctness of the above construction
is naturally split into checking several assertions:
namely, that $G'$ is an aligned LL($k$) grammar,
defines the same language as $G$,
and does not contain short rules.

\begin{claim}\label{Au_subset_A}
If a string $w$ is defined by a nonterminal $\nt{}{A}{u}$ in the new grammar,
then $w=yu$, where $y$ is defined by the nonterminal $A$ in the original grammar.
\end{claim}
\begin{proof}
Induction on the height of a parse tree for
$w$ as $\nt{}{A}{u}$.
\end{proof}

\begin{claim}
If a string $y$ is defined by a nonterminal $A$ in the original grammar,
then, in the new grammar, the nonterminal $\nt{}{A}{u}$ defines $yu$.
\end{claim}
\begin{proof}
Induction on the height of a parse tree for $y$ as $A$.
\end{proof}

The next claim establishes the correspondence
between parse trees in the original and the new grammar.

\begin{claim}\label{Bs_tree_to_B_tree}
Assume that there is a parse tree in $G'$,
wherein a $\nt{}{B}{s}$-subtree defines $ys$ by the rule $\nt{}{B}{s}\to\phi'$,
which was obtained from the rule $B\to\phi'$ in the original grammar,
and assume that the $\nt{}{B}{s}$-subtree
is followed by a string $z$.
Then, there is a parse tree in $G$,
with a $B$-subtree that defines $y$ by the rule $B\to\phi$,
and is followed by the string $sz$.

Furthermore, if $|s| < k-1$, then $z=\eps$.
\end{claim}
\begin{proof}
Induction on the depth of the $\nt{}{B}{s}$-subtree in the parse tree.
\end{proof}

\begin{figure}[t]
	\centering
	\includegraphics[scale=1]{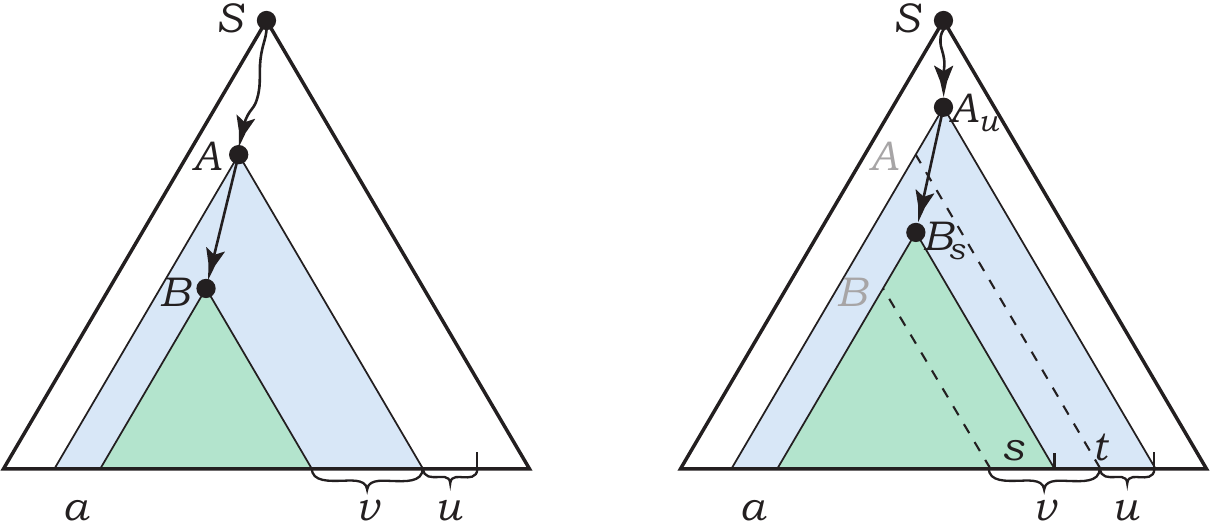}
	\caption{The rule $\nt{}{A}{u} \to \ldots\&\,a \nt{}{B}{s}t\,\&\ldots$ in $G'$
		is obtained from the rule $A \to\ldots\&\, a B v\,\&\ldots$ in $G$}
	\label{f:short_rule_elimination}
\end{figure}

Next, it is proved that $G'$ does not contain any short rules.

\begin{claim}
There are no short rules in $G'$.
\end{claim}
\begin{proof}
There are no short rules for nonterminals $\nt{}{A}{u}$ with $|u|=k-1$,
since, by Claim~\ref{Au_subset_A},
all strings defined by $\nt{}{A}{u}$ are of length as least $k-1$. 
 
And there are no short rules for nonterminals $\nt{}{A}{u}$ with $|u|<k-1$,
since, by Claim~\ref{Bs_tree_to_B_tree},
if $|u|<k-1$, then each $\nt{}{A}{u}$-subtree is followed by the empty string.
\end{proof}

Finally, it is proved that $G'$ is LL($k$).

\begin{claim}\label{conj_short_rules_ll_k_preserve}
Grammar $G'$ is LL($k$).
\end{claim}
\begin{proof}
Let $\tau_1'$ and $\tau_2'$ be parse trees in the grammar $G'$,
each containing an $\nt{}{A}{u}$-subtree,
and let the first $k$ leaves starting from the first leaves of subtrees 
$\nt{}{A}{u}$ form the same string $x$ in both trees.

\begin{figure}[t]
	\centering
	\includegraphics[scale=1]{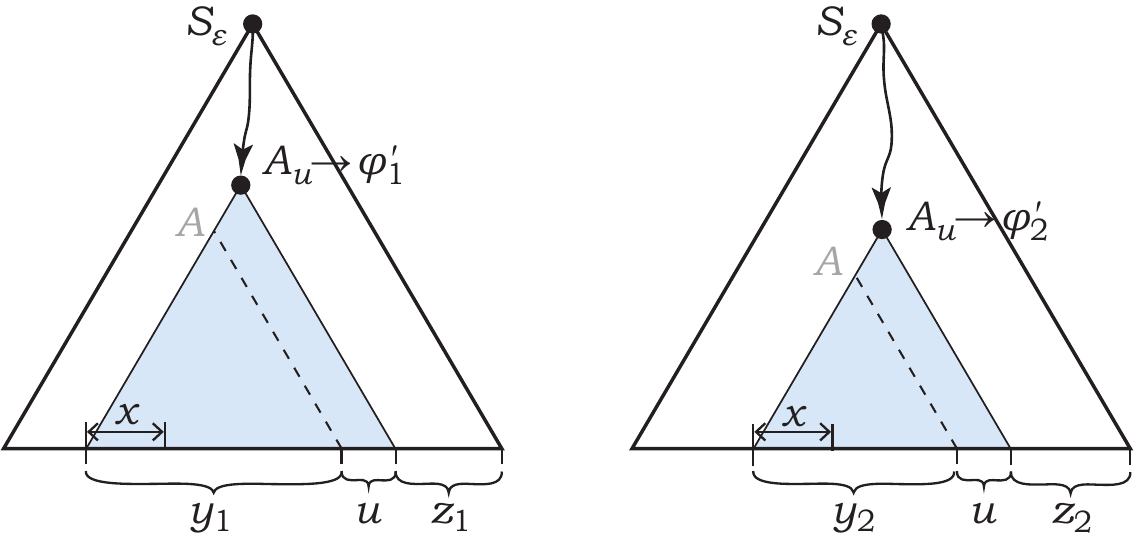}
	\caption{Parse trees $\tau_1'$ and$\tau_2'$ from Claim~\ref{conj_short_rules_ll_k_preserve}}
	\label{f:linllk_short_rules_LLk}
\end{figure}

For each $i\in\{1,2\}$,
let $y_iu$ be the string defined by the $\nt{}{A}{u}$-subtree in $\tau_i'$,
let $\nt{}{A}{u}\to\phi_i'$ be the rule applied to the root of this subtree,
and let $z_i$ be the string following this subtree,
as in Figure~\ref{f:linllk_short_rules_LLk}.

The rules $\nt{}{A}{u}\to\phi_1'$ and $\nt{}{A}{u}\to\phi_2'$
are obtained from some rules $A\to\phi_1$ and $A\to\phi_2$ of the original grammar.
By Claim~\ref{Bs_tree_to_B_tree} there exist parse trees
$\tau_1$ and $\tau_2$ in $G$,
each containing an $A$-subtree,
such that for each $i\in\{1,2\}$ the $A$-subtree in $\tau_i$
defines the string $y_i$ by the rule $A\to\phi_i$.

Then the first $k$ leaves of both parse trees,
starting with the first leaves of the $A$-subtrees,
form the same string $\first{y_1uz_1}=\first{y_2uz_2}=x$.

Since the grammar $G$ is LL($k$),
the rules used in $\tau_1$ and in $\tau_2$ coincide ($\phi_1=\phi_2$),
and hence the rules of $G'$ obtained from these rules coincide as well ($\phi_1'=\phi_2'$).
\end{proof}
Thus it has been shown
that $G'$ defines the same language as $G$,
is in LL($k$) and does not contain short rules.
Also $G'$ is aligned by the construction.
\end{proof}

\subsection{Reduction to LL(1)}

Once all short rules are eliminated from the grammar,
it can be further transformed to satisfy LL(1) property.

\begin{lemma}\label{conj_lookahead_reduction_lemma}
For each aligned LL($k$) grammar 
$G=(\Sigma, N, R, S)$ without short rules,
there exists an aligned LL(1) grammar $G'$
that defines the same language.
\end{lemma}
\begin{proof}
Nonterminals of the new grammar $G'=(\Sigma,N',R',\nt{\eps}{S}{})$
are of the form $\nt{u}{A}{}$, with $A \in N$ and $u \in \Sigma^{\leqslant k-1}$.

The intention is to have $\nt{u}{A}{}$ define strings from $L_G(A)$
with a prefix $u$ removed.
However,
the equality $L_{G'}(\nt{u}{A}{})=\set{x}{ux\in L_G(A)}$ generally does \textbf{not} hold,
but it holds that the string $ux$
is defined by a nonterminal $A$ \textit{inside some parse tree}
if and only if the string $x$ is defined by $\nt{u}{A}{}$ inside some parse tree.

The left subscript $u$ of a nonterminal $\nt{u}{A}{}$
works as buffer which stores the last $k-1$ symbols
read by parser.

The initial nonterminal of $G'$ is $\nt{\eps}{S}{}$,
which corresponds to $S$ with an empty buffer.

So $N'=\{\nt{u}{A}{}\;|\;A\in N,\,u\in\Sigma^{\leqslant k-1}\}$.
The rules of the new grammar $G'$ are separated in three sets:
$R_{buf}$,$R_G$ and $R_{empty}$.

Rules from $R_{buf}$ are responsible for filling the buffer. 
For each nonterminal $\nt{u}{A}{}$ 
with $|u|<k-1$ and for each symbol $b\in\Sigma$,
grammar $G'$ contains a rule attaching this symbol
to the buffer.
\begin{align*}
	\nt{u}{A}{} &\to b\,\nt{ub}{A}{}
\intertext{%
Rules from $R_G$ are used when the buffer is filled and thus
the parser can deduce which rule from the original grammar should be applied.
For each nonterminal $\nt{u}{A}{}\in N'$ and for each symbol $b \in \Sigma\cup\{\eps\}$,
where $|u|=k-1$ and $T(A, ub)$ is defined,
grammar $G'$ contains the rule obtained by \textit{removing} string $u$ from the rule $T(A,ub)$.
Suppose $T(A,ub)$ is of the form $A\to y$. Then, since $G$ does not contain short rules, $y=ux$ for some string $x\in\Sigma^*$ (note that short rules were eliminated exactly to make this part of construction work). Then the corresponding rule in $G'$ is
}
	\nt{u}{A}{} &\to x
\intertext{%
Now suppose $T(A,ub)$ is of the form $A\to aB^1 v_1\,\&\ldots\,\&\,aB^mv_m$.
Then $u$ should begin with $a$, and the corresponding rule in $G'$ is 
}
\nt{au'}{A}{} &\to \nt{u'}{B^1}{}v_1\,\&\ldots\,\&\,\nt{u'}{B^m}{}v_m, \quad \text{where $au'=u$}.
\intertext{%
Finally, rules from $R_{empty}$ are for the case when the buffer is not yet filled,
but the whole input string has already been consumed by the parser.
Namely, for each $\nt{u}{A}{}\in N'$,
with $|u| < k-1$ and with the entry $T(A, u)$ defined,
grammar $G'$ contains an empty rule.
}
	\nt{u}{A}{} &\to \eps
\end{align*}
Note that the sets $R_{buf},R_G,R_{empty}$ are disjoint.
For each rule $(\nt{u}{A}{}\to\phi')\in R_G$ it is always possible
to uniquely determine the rule $A\to\phi$ of the original grammar from which it was
obtained.
If $\nt{u}{A}{}\to \phi'$ is of the form $\nt{u}{A}{}\to x$ with $x\in\Sigma^*$, then $A\to\phi=A\to ux$, and if $\nt{u}{A}{}\to \phi'$ is of the form $\nt{au'}{A}{}\to \nt{u'}{B^1}{}v_1\,\&\ldots\,\&\,\nt{u'}{B^m}{}v_m$ then $A\to\phi=aB^1 v_1\,\&\ldots\,\&\,aB^mv_m$.

The proof that $G'$ is $LL(1)$
and defines the same language as $G$ is
given in a series of claims.

The correctness of the construction
is proved in the following three claims.

\begin{claim}\label{A_subset_uA}
Let $\nt{u}{A}{} \in N'$.
Let a parse tree of some string $w$ in $G$
contain an $A$-subtree that defines a string $ux$.
Then there exists a parse tree of $w$ in $G'$
that contains a $\nt{u}{A}{}$-subtree, which defines the string $x$.
\end{claim}
\begin{proof}
Induction on the height of the parse tree for $ux$ as $A$.
\end{proof}

\begin{claim}
If, in the grammar $G'$,
a nonterminal $\nt{u}{A}{}$ defines a string $x$,
then, in the original grammar, the nonterminal $A$ defines the string $ux$.
\end{claim}
\begin{proof}
Induction on the height of the parse tree for $x$ as $\nt{u}{A}{}$.
\end{proof}

\begin{claim}\label{uA_subset_A}
Assume that there is a parse tree in $G'$
with an $\nt{u}{A}{}$-subtree that defines a substring $x$
by the rule $\nt{u}{A}{}\to\phi'$,
and is followed by a string $z$.

Then, there exists a parse tree in $G$
with an $A$-subtree
that defines $ux$ and is followed by the string $z$.
Moreover, if the rule $\nt{u}{A}{}\to\phi'$ 
is obtained from the rule $A\to\phi$ of the original grammar
then this $A$-subtree defines $ux$ by the rule $A\to\phi$.
\end{claim}
\begin{proof}
Induction on the depth of the $\nt{u}{A}{}$-subtree.
\end{proof}

The grammar $G'$ is linear conjunctive by construction,
and Claims~\ref{uA_subset_A} and~\ref{A_subset_uA} together entail
$L(G')=L_{G'}(\nt{\eps}{S}{})=L_G(S)=L(G)$.
It remains to prove that $G'$ is $LL(1)$.

\begin{claim}
The grammar $G'$ is LL(1).
\end{claim}
\begin{proof}
Consider two parse trees $\tau_1'$ and $\tau_2'$ of the new grammar $G'$,
each containing an $\nt{u}{A}{}$-subtree,
and suppose that the strings starting from the first leaves of these subtrees either both begin with the same symbol or are both empty.
Denote this symbol as $b$ (if both strings are empty then $b=\eps$).

For each $i\in\{1,2\}$,
let $x_i$ be the string defined by the $\nt{u}{A}{}$-subtree in $\tau_i'$,
let $\nt{u}{A}{}\to\phi_i'$ be the rule applied to its root, 
and let $z_i$ be the string following the subtree s,
that $b=\first[1]{x_iz_i}$
Now it will be proved that $\phi_1'=\phi_2'$.

The proof is given separately for nonterminals $\nt{u}{A}{}$ with $|u|<k-1$,
and for nonterminals $\nt{u}{A}{}$ with $|u|=k-1$.
First, let $|u|<k-1$.
Then, each of the rules $\nt{u}{A}{}\to\phi_1'$ and $\nt{u}{A}{}\to\phi_2'$
is either in $R_{buf}$ or in $R_{empty}$.
Consider the cases.
\begin{itemize}
\item
If both rules are in $R_{empty}$, then $\phi_1'=\phi_2'=\eps$.
\item
If both rules are in $R_{buf}$, then $\phi_1'=\phi_2'=b\,\nt{ub}{A}{}$.
\item
Suppose that one of the rules, say $\phi_1$, is in $R_{buf}$,
and the other is in $R_{empty}$.
Then $\phi_1'=b\,\nt{ub}{A}{}$ and $\phi_2'=\eps$. 
Hence $b\neq\eps$, because $\phi_1'\in R_{buf}$.

On the other hand, since $\phi_2'\in R_{empty}$,
then $\eps\in L_{G'}(\nt{u}{A}{})$,
and, by Claim~\ref{uA_subset_A}, there is a parse tree in $G$
with an $A$-subtree that defines the string $ux_2=u$,
and the leaves to the right of the subtree form the string $z_2$.
The grammar $G$ does not contain short rules,
and hence $|u|<k-1$ entails
$z_2=\eps$, and therefore $b=\first[1]{x_2z_2}=\eps$.
The contradiction obtained implies that this case is actually impossible.
\end{itemize}

Now suppose that $|u|=k-1$.
Then both rules $\nt{u}{A}{}\to\phi_1'$ and $\nt{u}{A}{}\to\phi_2'$ are in $R_G$,
and therefore are obtained from some rules $A\to\phi_1$ and $A\to\phi_2$
in the original grammar.

By Claim~\ref{uA_subset_A}, there are parse trees $\tau_1$ and $\tau_2$ in $G$,
such that for each $i\in\{1,2\}$, the parse tree $\tau_i$
contains an $A$-subtree that defines the string $ux_i$,
the rule applied to the root is $A\to\phi_i$,
and the leaves to the right of the subtree form the string $z_i$.

Then the first $k$ leaves of these parse trees,
starting with the first leaves of $A$-subtrees,
form the same string $\first{ux_1z_1}=\first{ux_2z_2}=ub$.

Since $G$ is LL($k$), this is the same rule
($\phi_1=\phi_2$),
and hence $\phi_1'=\phi_2'$.
\qedhere
\end{proof}

Now it has been proved that $G'$ is an LL(1) linear conjunctive grammar
that defines the same language as $G$. Then by Lemma~\ref{for_each_grammar_exists_aligned_grammar}
there exists an aligned LL(1) grammar which defines the same language as $G$,
and therefore the proof of Lemma~\ref{conj_lookahead_reduction_lemma} is complete.
\end{proof}

Together, Lemmata~\ref{conj_short_rule_elimination_lemma}
and~\ref{conj_lookahead_reduction_lemma}
constitute the proof of Theorem~\ref{conj_ll_k_to_ll_1_theorem}.

\section{An efficient parser for aligned LL(1) linear conjunctive grammars}\label{parser_section}

A \emph{parser} for a grammar $G$ is an algorithm
that decides whether a given string $w\in \Sigma^*$
is defined by the grammar.
For an ordinary \textbf{LL($k$) grammar} without conjunction,
there exists a canonical parser
that attempts to reconstruct a parse tree for the input string,
while reading it from left to right,
At each step the parser uses the next $k$ input symbols to determine, which rule to apply.
The parser uses stack memory, which contains a string of symbols from $\Sigma\cup N$
representing the projected form of the remaining input string~\cite{Knuth_LL,Kurkisuonio,RosenkrantzStearns}.

A classical LL($k$) parser can be generalized to LL($k$) conjunctive grammars,
but the generalized parser, instead of a stack,
requires a more complicated data structure:
a \emph{tree-structured stack}, which contains multiple top symbols
and a single bottom~\cite{AizikowitzKaminski,AizikowitzKaminski_linear,ConjunctiveLL,BooleanLL}.

In this section it is shown
that in the case of LL($k$) linear conjunctive grammars,
instead of a complicated tree-structured stack,
it is sufficient to use a set of standard stacks.
Moreover, it will be proved that the number of stacks in the set
never exceeds the number of nonterminal symbols in the grammar
(Lemma~\ref{profiles_are_small}),
and this fact will allow an implementation of this parser
that uses logarithmic space (Theorem~\ref{conj_in_logspace}).

Let $G=(\Sigma,N,R,S)$
be an LL($k$) linear conjunctive grammar.
By Lemmata~\ref{conj_ll_k_to_ll_1_theorem} and~\ref{for_each_grammar_exists_aligned_grammar},
it may be assumed that $G$ is aligned and LL(1).
Let $w=a_1\ldots a_n$ be an input string.
At each step of the computation,
the parser's configuration is a pair $(Z,a_i \cdots a_n)$,
where $Z$ is a set of conjuncts of the form $\{A_1v_1, \ldots, A_k v_k\}$,
called a \textit{stack set},
and $a_i \cdots a_n$ is an unread suffix of the input string.
The following invariant is maintained:
the entire input string $w$ is defined by the grammar
if and only if
the unread suffix $a_i\cdots a_n$ is defined by each conjunct in $Z$,
that is, $a_i \cdots a_n \in L_G(A v)$ for each $Av \in Z$.

The parser's initial configuration is a pair $(\{S\},w)$:
there is a single stack containing $S$,
and the whole input remains unread.

At each step of its computation,
the parser reads the next input symbol
and processes each conjunct in its stack set
according to this symbol and the LL(1) table.
Let $(\{A_1v_1,\ldots,A_k v_k\},a_i \ldots a_n)$ be the current parser's configuration.
Let $a=a_i$ be the next input symbol
(if the whole input is already consumed, then $a=\eps$).
Then, for each conjunct $A_j v_j$,
the parser determines the correct rule for $A_j$
and substitutes it for $A_j$ as follows.
\begin{itemize}
\item
If $T(A_j,a)$ is not defined,
then the parser reports a parse error and halts.
\item
If $T(A_j,a)=A_j\to y_j$,
then the parser checks that the unread suffix $a_i\cdots a_n$ of the input
coincides with $y_jv_j$.
If this is the case,
then the parser removes the conjunct $A_j v_j$ from the stack set,
otherwise it reports a parsing error and halts.
\item
If $T(A_j,a)=A_j\to aB_{j,1}v_{j,1}\,\&\ldots\&\,aB_{j,m_j}v_{j,m_j}$,
then the parser replaces each conjunct $A_jv_j$ from the stack set
with the set of conjuncts $\{B_{j,1}v_{j,1}v_j,\ldots, B_{j,m_j}v_{j,m_j}v_j\}$.
\end{itemize}

Assume that the conjuncts in the stack set are enumerated,
so that the rules $T(A_j,a)$
applied to the first $r$ conjuncts contain nonterminals,
while the rules for the remaining conjuncts
$A_{r+1}v_{r+1}, \ldots, A_k v_k$
are of the form $T(A_j,a)=A_j\to y_j$.
Altogether, the following rules are used.
\begin{align*}
T(A_1,a) &= A_1\to aB_{1,1}v_{1,1}\,\&\ldots\&\,aB_{1,m_1}v_{1,m_1}\\
&\;\vdots\\
T(A_r,a) &= A_r\to aB_{r,1}v_{r,1}\,\&\ldots\&\,aB_{r,m_r}v_{r,m_r}\\
\vspace{2cm}\\
T(A_{r+1},a) &= A_{r+1}\to y_{r+1};\quad y_{r+1}v_{r+1}=a_i\cdots a_n\\
&\;\vdots\\
T(A_k,a) &=A_k\to y_k,
    && \text{where } y_kv_k=a_i\cdots a_n
\end{align*}
Using these rules, the computation step proceeds as follows.
\begin{align*}
(\{A_1v_1,\ldots,A_kv_k\},a_i a_{i+1}\cdots a_n)
\rightarrow
(\{&B_{1,1}v_{1,1}v_1,\ldots, B_{1,m_1}v_{1,m_1}v_1,\\
&\quad\quad\vdots\\
&B_{r,1}v_{r,1}v_r,\ldots, B_{r,m_r}v_{r,m_r}v_r\},a_{i+1}\cdots a_n)
\end{align*}

If, at some step, the stack set happens to be empty,
then the parser has actually already verified that the string is defined by the grammar.
At the remaining steps, it switches to ``idle mode''
and reads the rest of the input symbols.

Since the parser reads one input symbol at each step,
if the computation goes successfully, 
the parser reaches the configuration $(Z_n,\eps)$
after exactly $n=|w|$ steps.
Then, at the last $(n+1)$-st step,
the parser tries to apply to each conjunct $Av\in Z_n$
the rule $T(A,\eps)$, which can only be of the form $A \to \epsilon$.
If all these rules exist,
the parser completes this last step in the configuration $(Z_{n+1},\epsilon)$.
If $Z_{n+1}=\emptyset$,
then the computation is accepting.
If either $Z_{n+1}\neq\emptyset$, or the computation halted earlier,
then the computation is rejecting.

Thus, the computation consists of exactly $n+1$ steps.
At each step of the computation,
rules are applied to each element from the stack set,
and the next input symbol is read
(at the last step, no symbol is read).
As a result of rule application,
conjuncts ``spawn'', that is,
are substituted with a (possibly empty) set of new conjuncts.
Note that the total number of \textit{different} conjuncts may decrease
both because some old conjuncts have no descendants,
and because some new conjuncts coincide.

Consider an accepting computation of the parser.
\begin{equation*}
    (\{S\},w)=(Z_0,a_1\cdots a_n)\rightarrow(Z_1,a_2\cdots a_n)\rightarrow\ldots\rightarrow (Z_{n+1},\eps)=(\emptyset,\eps)
\end{equation*}
Each conjunct from $Z_{i+1}$ is a descendant
of a conjunct from the previous stack set $Z_i$.
Formally, the notion of a \emph{descendant} is defined as follows.

Let $\alpha\in Z_i$ be any conjunct.
The sets $Z^{\alpha,i}_j$, with $j\in\{i,\ldots,n+1 \}$ and $Z^{\alpha,i}_j\subseteq Z_j$,
are constructed inductively as follows.

For $j=i$, let $Z^{\alpha,i}_i=\{\alpha\}$.
Now let us define the set $Z^{\alpha,i}_{j+1}$
using the already constructed set $Z^{\alpha,i}_j$.

Let $j>i$, $Z^{\alpha,i}_{j-1}=\{A_1v_1,\ldots,A_kv_k\}$,
and let $a=a_i$ be the next symbol of the input string
(if the whole input is already read, then $a=\eps$).

Each conjunct $A_p v_p\in Z^{\alpha,i}_{j-1}$
gives rise to the set $Next(A_p v_p)\subseteq Z_j$, which is defined as follows.
If $T(A_p,a)=A_p\to y$, then $Next(A_p v_p)=\emptyset$.
If $T(A_p,a)=A_p\to aB_{p,1}v_{p,1}\,\&\ldots\&\,aB_{p,m_p}v_{p,m_p}$,
then $Next(A_pv_p)=\{B_{p,1}v_{p,1}v_p,\ldots, B_{p,m_p}v_{p,m_p}v_p\}$.

The set $Z^{\alpha,i}_j$ is then defined as $\bigcup_{p=1}^k Next(A_pv_p)$.

All conjuncts from the sets $Z^{\alpha,i}_i,\ldots,Z^{\alpha,i}_{n+1}$ are called \textit{descendants} of $\alpha\in Z_i$. 

Note, that since each conjunct from $Z_j$
is a descendant of some conjunct from $Z_{j-1}$,
the stack set $Z_j$ at the $j$-th configuration
equals $\bigcup_{\alpha\in Z_{j-1}} Z^{\alpha,j-1}_j$.
However, some conjuncts from $Z_j$ can at the same time
be descendants of several conjuncts from $Z_{j-1}$,
so that the sets $Z^{\alpha,j-1}_j$ can intersect for different $\alpha$. 

Each conjunct is a descendant of the conjunct $S$ from the initial configuration,
thus $Z_j=Z^{S,0}_j$.
By the time the computation ends,
each conjunct disappears from the stack set,
hence, for all $\alpha$ and for all $j$,
the set $Z^{\alpha,j}_{n+1}$ is empty.

Now let us check the correctness of the above parsing algorithm,
and also establish a correspondence between parse trees and accepting computations.

The next lemma states that each accepting computation on some string
corresponds to a parse tree of that string.
 
\begin{lemma}\label{calculation_to_tree}
Let $G$ be an aligned LL(1) linear conjunctive grammar, and let
$(\{S\},w)=(Z_0,a_1\cdots a_n)\rightarrow(Z_1,a_2\cdots a_n)\rightarrow\ldots\rightarrow(Z_{n+1},\eps)=(\emptyset,\eps)$
be the accepting computation of the $G$-parser on the string $w$.
Then there exists a parse tree $\tau$ for $w$,
and, for each $i\in\{0,\ldots, n\}$, $A\in N$ and $v\in\Sigma^*$,
the next two statements are equivalent:
\begin{itemize}
\item There exists an $A$-subtree in $\tau$, such that
the leaves to the left of the subtree form the string $a_1\cdots a_i$,
while the leaves to the right of the subtree form the string $v$.
\item The stack set $Z_i$ contains the conjunct $Av$.
\end{itemize}
\end{lemma}
\begin{proof}
The proof introduces some notation
for fragments of a parser's computation
evolving from a single conjunct occurring at some $i$-th step,
and comprised of all its descendants.
This is a kind of subcomputation that ignores all conjuncts
other than the descendants of a chosen conjunct.

Let $\alpha=Av$ be a conjunct in $Z_i$.
Then, an \emph{$(\alpha,i)$-generated computation}
is defined as a sequence
$(Z^{\alpha,i}_i,a_{i+1}\cdots a_n),(Z^{\alpha,i}_{i+1},a_{i+1}\cdots a_n),\ldots,(Z^{\alpha,i}_\ell,a_{\ell+1}\cdots a_n)=(\emptyset,a_{\ell+1}\cdots a_n)$,
where $\ell$ is the first index of configuration, in which $\alpha$ has no descendants.

By induction on the length of $(\alpha,i)$-generated computation
(that is, on $\ell-i$), it is proved that:
\begin{enumerate}
\item\label{descendants_end_with_v}
Each descendant of $\alpha$ ends with $v$,
and hence is of the form $u'A'v'v$.
\item\label{descendants_subtrees}
Let $y$ be a string, such that $yv=a_{i+1}\cdots a_n$.
Then, there exists a parse tree $\tau_\alpha$ for $y$,
with its root labelled with $A$,
such that, for each $j\in\{i,\ldots,\ell\}$,
the next two statements are equivalent:
\begin{itemize}
    \item There exists an $A'$-subtree in $\tau_\alpha$,
    such that the leaves to the left of the subtree
    form the string $u'=a_{i+1}\ldots a_j$,
    and the leaves to the right of the subtree form the string $v'$.
    \item
    The stack set $Z^{(\alpha,i)}_j$ contains the conjunct $A'v'v$.
\end{itemize}
\end{enumerate}

Note, that for the conjunct $S$ from the initial configuration,
point~\ref{descendants_end_with_v} is trivially satisfied since $v=\eps$,
and point~\ref{descendants_subtrees} is exactly the statement of the lemma.

In the base case of the induction,
the length of a $(\alpha,i)$-generated computation is 1,
and at the $i$-th step
the parser applies a rule $A\to y$ to the conjunct $Av$.
Then, since the computation is accepting,
it must hold that $a_{i+1}\cdots a_n=yv$.
Then the required parse tree for $y$
consists of a single rule $A\to y$ applied to the root.

Now assume that at the $i$-th step
the parser applies
a rule $A\to aB_1v_1\,\&\ldots\&\,a B_m v_m$
to the conjunct $Av$,
with $a=a_i$.
Then, $B_1v_1v$,\ldots,$B_mv_mv$ are all the descendants of $Av$
from the $i$-th configuration.
By the induction hypothesis,
for each conjunct $B_jv_jv$, there exists a string $z_j$,
such that $z_jv_jv=a_{i+1}\cdots a_n$,
all descendants of $B_jv_jv$ end with $v_jv$,
and there exists a parse tree of $z_j$
with its root labelled with $B_j$, as in point~\ref{descendants_subtrees}.

Therefore, $a_i\cdots a_n=yv$
for some string $y=az_1v_1=\ldots=az_mv_m$,
and all descendants of $Av$ end with $v$.

Now, parse trees for the strings $z_1,\ldots,z_j$
with the roots $B_1,\ldots, B_j$
can be merged into one parse tree for $y$
with the root $A$,
in which the rule $A\to aB_1v_1\,\&\ldots\&\,a B_m v_m$ is applied to the root,
\end{proof}

The next lemma states, that each parse tree corresponds to an accepting computation.

\begin{lemma}
Let $G$ be an aligned LL(1) grammar,
and let $\tau$ be a parse tree for a string $w$.
Then there exists a (unique) accepting computation
$(\{S\},w)=(Z_0,a_1\cdots a_n),\ldots,(Z_{n+1},\eps)=(\emptyset,\eps)$,
and, for all $i\in\{0,\ldots,n\}$, $A\in N$ and $v\in\Sigma^*$,
the following statements are equivalent:
\begin{itemize}
\item The stack set $Z_i$ contains the conjunct $Av$.
\item The parse tree $\tau$ contains an $A$-subtree,
with the leaves to the right of the subtree forming the string $v$.
\end{itemize}
\end{lemma}
\begin{proof}
Let us inductively construct configurations
$(Z_0,a_1\cdots a_n),\ldots,(Z_{n+1},\eps)$ of the accepting computation,
at each step assuming, that the last constructed configuration
satisfies the condition in the lemma.

The base case $i=0$ corresponds to the initial configuration $(\{S\},w)$.
The conjunct $S\in Z_0$ in this case corresponds to the whole tree $\tau$. 

Now suppose that $0<i\leqslant n+1$,
and the configuration $(Z_{i-1},t_{i-1})$ is already defined.

To define the next configuration $(Z_i,a_{i+1}\cdots a_n)$,
it is sufficient to show, that, for each conjunct $Av\in Z_{i-1}$,
the rule $T(A,a_i)$ is defined, and that,
if that rule is of the form $T(A,a_i)=A\to y$, then $yv=a_i\ldots a_n$.

Consider any conjunct $Av\in Z_{i-1}$.
By the induction hypothesis, there exists an $A$-subtree in $\tau$,
such that the leaves to the left of the subtree
form the string $a_1\ldots a_{i-1}$,
while leaves to the right of the subtree form the string $v$.
Denote that subtree by $\tau_A$.
Since the grammar $G$ is LL(1),
the rule applied to the root of $\tau_A$ is $T(A,a)$,
where $a=a_i$ ($a=\eps$ if $i=n+1$).
In particular, $T(A,a)$ is defined.
If $T(A,a)=A\to y$, then the existence of $\tau_A$ implies $yv=a_i\cdots a_n$.

Now assume the rule applied to $A$ is $T(A,a)=A\to aB_1v_1\,\&\ldots\&\,aB_mv_m$.
Then, $\tau$ contains subtrees with roots $B_1,\ldots,B_m$,
and, for each $j\in\{1,\ldots,m\}$,
the leaves to the left of the $B_j$-subtree
form the string $a_1\ldots a_i$,
while the leaves to the right of the $B_j$-subtree
form the string $v_jv$.

At the $i$-th step,
the parser has to apply the same rule $A\to aB_1v_1\,\&\ldots\&\,aB_mv_m$
to the conjunct $Av$,
and therefore conjunct $Av$ gives rise
to the set of descendants $Z^{Av,i-1}_i=\{B_1v_1v,\ldots,B_mv_mv\}$.
For every such descendant $B_jv_jv$, as it was mentioned,
there is a subtree in $\tau$,
such that the leaves to the left of that subtree
form the string $a_1\cdots a_i$,
while the leaves to the right of the subtree form the string $v_jv$.

Therefore, the parser is able to perform the $i$-th step of the computation
and update its configuration to $(Z_i,t_i)$,
where $Z_i=\bigcup_{Av\in Z_{i-1}} Z^{Av,i-1}_i$.
Each conjunct $B_jv_jv\in Z_i$ is a descendant of some conjunct $Av\in Z_{i-1}$,
hence, for each conjunct, there exists a subtree from the statement of the lemma.

Vice versa, assume that $\tau$ contains a $B$-subtree $\tau_B$,
with the leaves to the left of $\tau_B$ forming the string $a_1\ldots a_i$,
and with the leaves to the right of $\tau_B$ forming the string $v''$.
Let $A$ be the nonterminal labelling the immediate ancestor of $\tau_B$,
and let $v$ be the string following the $A$-subtree,
so that $v''=v'v$, for some string $v'$.
Since the grammar is aligned,
the leaves to the left of the $A$-subtree
form the string $a_1\ldots a_{i-1}$,
and, by the induction hypothesis,
the stack set $Z_{i-1}$ contains a conjunct $Av$
corresponding to the $A$-subtree.
The rule applied to the root of the $A$-subtree
and the rule applied to the conjunct $Av$ are both $T(A,a)$,
thus conjuncts from $Z^{Av,i-1}_i$ one-to-one correspond
to the immediate descendants of $A$.
Therefore, the stack set $Z_i$
contains a conjunct $Bv'v\in Next(Av,a)$ corresponding to $\tau_B$. 

Thus, the sequence of configurations
$(Z_0,t_0),(Z_1,t_1),\ldots, (Z_n,\eps), (Z_{n+1},\eps)$ has been defined,
and it remains to show that $Z_{n+1}$ is empty.

By construction, if $A_n$ contains a conjunct $Av$,
then there is a subtree in $\tau$,
such that the leaves to the left of the subtree
form the whole input string $w$,
while the leaves to the right of the subtree form the string $v$.
Then, of course, $v=\eps$, and, since the grammar is aligned, $T(A,\eps)=A\to\eps$.

Therefore, $Z_n$ can contain only conjuncts
consisting of a single nonterminal,
and at the $n$-th step the parser
is able to apply an empty rule to each of these nonterminals.
Hence, $Z_{n+1}=\emptyset$,
and thus the computation $(Z_0,t_0),(Z_1,t_1),\ldots, (Z_{n+1},\eps)$ is accepting.
By construction, it satisfies the statement of the lemma.
\end{proof}

Finally, it is possible to prove
the main property of the described parser,
which implies its efficiency:
the size of the stack set
is bounded by the number of nonterminals in the grammar.

\begin{lemma}\label{profiles_are_small}
Let $G$ be an aligned LL(1) grammar, $w\in L(G)$,
and let $s$ be a prefix of $w$.
Assume that the parser's stack set after reading the prefix $s$
is $Z=\{A_1v_1,\ldots,A_kv_k\}$.
Then, for every two elements $A_1v_1,A_2v_2\in Z$,
it holds that $A_1=A_2\Rightarrow v_1=v_2$, and therefore $|Z|\leqslant |N|$.
\end{lemma}
\begin{proof}
By Lemma~\ref{calculation_to_tree},
there exists a parse tree $\tau$ for $w$,
such that, for each conjunct $Av_i$, with $i\in\{1,2\}$,
there exists an $A$-subtree
with the leaves to the left of it forming the string $s$,
and with the leaves to the right of the subtree forming the string $v_i$.
Then, by Lemma~\ref{LL_subtrees_equivalence},
both subtrees define the same string $y$.
Hence $w=yv_1=yv_2$, and therefore $v_1=v_2$.
\end{proof}

\section{Parsing in LOGSPACE}\label{conj_in_logspace_section}

Lemma~\ref{profiles_are_small} proved in the previous section
makes it possible to develop an improved implementation of a parser,
which uses logarithmic space and still works in linear time.

\begin{theorem}\label{conj_in_logspace}
The language defined by each LL($k$) linear conjunctive grammar $G=(\Sigma,N,R,S)$
is decidable in logarithmic space and linear time.
\end{theorem}
\begin{proof}
By Theorem~\ref{conj_ll_k_to_ll_1_theorem},
there exists an aligned LL(1) linear conjunctive grammar $G'$
that defines the same language as $G$,
hence it can be assumed that $G$ is aligned and LL(1).

Consider the LL(1)-parser for $G$
described in Section~\ref{parser_section}.
Its data structures shall now be revised.

By definition, a configuration of an LL(1)-parser at each step of a computation
is a pair $(Z,a_i\cdots a_n)$,
where $Z$ is a stack set of the form $\{A_1v_1,\ldots,A_kv_k\}$,
and $a_i\cdots a_n$ is the unread suffix of the input string.
 
Instead of ``tails'' $v_1,\ldots,v_k$,
the logspace-parser stores only their lengths.
Therefore, each conjunct $Av$ is encoded in the logspace-parser
as the pair $(A,|v|))$.
Instead of the suffix $a_i\cdots a_n$,
the logspace-parser stores only the current position $i$.

Therefore, the corresponding configuration of the logspace-parser
is a pair $(Z',i)$,
where $Z'=\{(A_1,|v_1|),\ldots, (A_k,|v_k|)\}$.

Let $a=a_i$ be the next symbol of the input.
At the $i$-th step of the computation,
the LL(1) parser applies the rule $T(A_j,a)$ to each conjunct $A_jv_j$.
The logspace-parser implements this in the following way.

If $T(A_j,a)=A_j\to y$, with $y\in\Sigma^*$,
then the logspace-parser checks that the substring $a_i\cdots a_{i+|y|-1}$
coincides with $y$.
Note that since $|y|$ is bounded by the size of the grammar,
this is done in constant time.
If the strings indeed coincide,
then the logspace-parser just removes the pair $(A_j,|v_j|)$ from the stack set,
and otherwise the logspace-parser reports a parse error.

If $T(A_j,a)=A_j\to aB_{j,1}v_{j,1}\&\ldots\&aB_{j,p_j}v_{j,p_j}$,
then the logspace-parser checks
that each of the strings $v_{j,1},\ldots,v_{j,p_j}$
coincides with the corresponding substring of the input string,
and replaces each pair $(A_j,|v_j|)$ with the set $\{(a,B_{j,1},|v_j|+|v_{j,1}|),\ldots,(a,B_{j,p_j},|v_j|+|v_{j,p_j}|)\}$.
This is also done in constant time,
since all $p_j$ and $|v_{j,i}|$ are bounded by the size of the grammar.
\end{proof}


\end{document}